# Holographic Amplitude-Modulated (AM) Leaky-Wave Antennas for Near-Field and Far-Field Applications

Geng-Bo Wu, *Member, IEEE*, Ka Fai Chan, *Member, IEEE* and Chi Hou Chan, *Fellow, IEEE*

*Abstract*—Amplitude-modulated (AM) leaky-wave antenna (LWA), a concept following amplitude modulation technique from classical communications theory, is a promising structure that enables transforming traveling wave into the radiating wave. In this paper, we provide a different perspective based on the classical holographic theory to gain insight into the physical mechanism of AM LWA and design novel LWAs. In analogy to the classical optical Gabor hologram, we demonstrate that only the *amplitude* variation of the traveling wave is needed to record both the amplitude and phase information of the object wave. The consistency between the holography theory and previous spatial spectrum approach for explaining AM LWA operating mechanism is also demonstrated. For validation purpose, two novel millimeter-wave (mmW) holographic AM LWAs based on the substrate integrated inset dielectric waveguide (IDW) are designed. The first one is for far-field high-gain applications while the second is for near-field focusing (NFF) applications. Both simulated and measured results demonstrate the effectiveness of the AM holography theory for AM LWAs analysis and design.

## I. INTRODUCTION

Leaky-wave antennas (LWAs), supporting traveling waves with progressive leakage along the wave-guiding structure, are a promising candidate for millimeter-wave (mmW) high-gain antenna applications [1]-[2]. Low profile, simple feeding mechanism, and inherent frequency beam scanning capability are the several advantages of LWA in comparison to other high-gain solutions such as constrained-fed array antennas [3] and space-fed reflector/lens antennas [4]-[7]. LWAs can find widespread applications in imaging [8], automotive radar [9], and direction of arrival estimation [10], etc.

The holography concept, originating from optics, can be extended to the RF band, offering new potential for antenna design. The radiation field from the spatial feed was first used as the reference wave for holographic antenna design [11]. Afterward, surface wave-fed holographic antenna, i.e., holographic LWA, was proposed with advantages of simplified feeding and antenna structures [12]-[14]. The design of a periodic LWA can be interpreted as adding the scatterers/perturbations at the minimal intensity positions of the interference pattern between the incident surface wave and the plane wave coming from the direction of maximum radiation. By again exciting the reference wave, the wave-guiding structure with scatterers/perturbations can reconstruct the object plane wave radiating in the desired direction. Alternatively, Sievenpiper *et al*. proposed that the interference pattern can also be characterized by the surface impedance [15]-[16], providing another effective approach for sinusoidally modulated reactance surfaces (SMRS) and modulated metasurfaces (MTS) antenna design besides local modulation of surface impedance [17]-[19]. The amplitude holography theory has been used to design reconfigurable metamaterial antennas in [20]-[22] for far-field applications. Nevertheless, the resonant unit cell/scatterers in metamaterial antennas will inherently introduce phase shifting to the carrier wave, making the amplitude holography just a zeroth-order approximation for reconfigurable metamaterial antennas design.

A new amplitude-modulated (AM) LWA concept, inspired by the well-known amplitude modulation technique in classical communications theory, was recently proposed by Wu *et al* [23]. Compared to periodic LWAs and SMRS, the sinusoidal AM LWAs feature the advantage of free of any undesired higher-order Floquet modes ($|n| \geq 2$). Moreover, since the surface impedance is uniform for AM LWAs, the time-consuming eigenmode simulation in the SMRS design process can be eliminated.

In this article, the classical optical Gabor holography is developed to reinterpret the physical mechanism of the AM LWA, which allows a more in-depth understanding of AM LWAs and enables novel AM LWAs design. It is demonstrated that the modulating waveform in [23] can be viewed as the intensity interference pattern between the reference and object waves. Two novel mmW holographic AM LWAs based on substrate integrated inset dielectric waveguide (IDW) is subsequently designed for demonstration purpose. The first design is for high-gain far-field applications, in which the object wave is a plane wave. The amplitude holography theory

This work was supported by the Hong Kong Research Grants Council under Grant T42-103/16-N and CityU 11250216.

G. B. Wu and C. H. Chan are with the State Key Laboratory of Terahertz and Millimeter Waves, and also with Department of Electrical Engineering, City University of Hong Kong (CityU), Hong Kong, China (corresponding to: eechic@cityu.edu.hk ).

K. F. Chan is with the State Key Laboratory of Terahertz and Millimeter Waves, City University of Hong Kong (CityU), Hong Kong, China.

is further utilized to synthesize a near-field focusing (NFF) LWA, which has never been reported in the open literature to the best of our knowledge. Modulating the width of the IDW never tunes the phase of the carrier wave. Therefore, the amplitude holography theory is more accurate for the IDW LWA design compared to metamaterial antennas [20]-[22]. Both numerical and experimental results, in good agreement, are given to verify the holographic AM LWA designs.

## II. HOLOGRAPHIC AM LWA THEORY

It is known that an antenna should provide proper aperture amplitude and phase distributions related to the desired far-field radiation pattern. The aperture phase distribution of an antenna is apparently important because it directly determines the propagation direction of radiating EM waves. However, the AM LWA [23] is a type of spatial amplitude-only modulation without modulating the phase angle of the traveling wave. The aperture phase distribution of an AM LWA is a simple progressive phase front, which is identical to that of the traditional unmodulated wave-guiding structure. Therefore, one may raise a legitimate question: AM LWAs, as an amplitude-only variation structure without phase modulation, how to generate the desired object wave with both amplitude and phase information?

In fact, the exactly same problem is encountered for image recording in the optical community. The object image that needs to be recorded contains both amplitude and phase information. The available recording media such as photographic emulsions, however, are only responsive to the intensity of the light wave. In 1948, Gabor proposed a seminal imaging process, known as holography, to convert the phase information of waves into intensity variations [24]. The holography technology uses a known reference wave $E_r$ to interfere with the desired object wave $E_o$ and photographically record only the amplitude information of their interference pattern in the recording material. By simply again illuminating the amplitude-only hologram with the reference wave, the recorded information can be decoded and the object wave can be reconstructed.

In this section, we answer the above-mentioned complex amplitude information-carrying problem for AM LWAs based on the classical Gabor holography theory. The holography generally involves the two processes of recording and reconstruction.

### A. Recording

Suppose a one-dimensional (1-D) AM LWA is designed to generate a far-field high-gain beam with a radiation angle $\theta_b$ with respect to the broadside direction as shown in Fig. 1(a). The corresponding object wave in this case is a plane wave and can be written as:
$$E_o = A_o \exp[-j(k_0 \sin\theta_b x + k_0 \cos\theta_b z)] \quad (1)$$
where $A_o$ is the amplitude constant and $k_0$ is the free-space

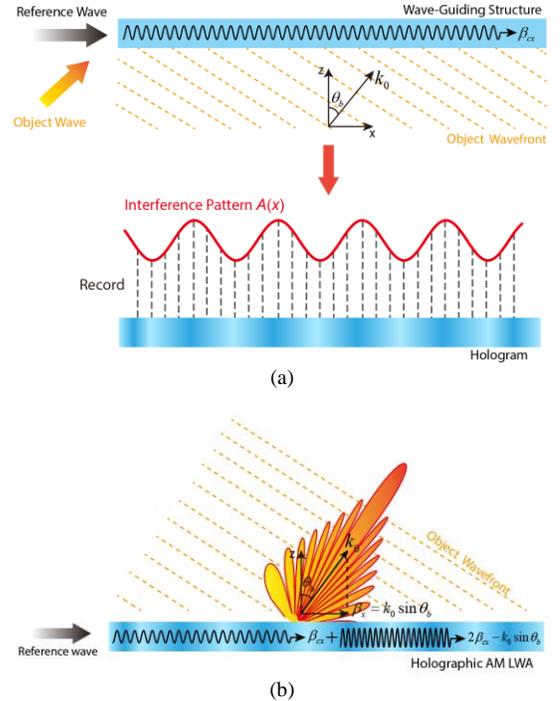

Fig. 1. Concept of the holographic AM LWA: (a) Recording process; (b) Reconstruction process.

wavenumber. The unmodulated slow surface wave along the wave-guiding structure is used as the reference wave:
$$E_r = A_r \exp(-j\beta_{cx} x) \quad (2)$$
where $A_r$ is a positive constant for a uniform traveling-wave structure. A slow-wave transmission line is assumed here, and thereby $\beta_{cx} > k_0$.

In analogy to the Gabor hologram, it is the intensity of the interference pattern between the object wave and reference wave in the antenna aperture plane that is recorded in the wave-guiding structure. The amplitude variation/envelop of the traveling wave along the wave-guiding structure can be written as:
$$A(x) = |E_o + E_r|^2$$
$$= |E_o|^2 + |E_r|^2 + E_r^* E_o + E_r E_o^* \quad (3)$$

Substituting (1) and (2) into (3), we obtain
$$A(x) = A_o^2 + A_r^2 + 2A_r A_o \cos[\arg(E_r) - \arg(E_o)]$$
$$= A_o^2 + A_r^2 + 2A_r A_o \cos(\beta_{cx} x - k_0 \sin\theta_b x) \quad (4)$$

The first two terms on the right-hand side of (4) depend only on the intensities of the individual waves, and the third term depends on their relative phase. Although the intensity of interference pattern $A(x)$ is a positive real number, the complete information involving both the amplitude $A_r$ and phase information $-\beta_{cx} x$ of the object wave $E_o$ is coded in the third term of (4). In this manner, the phase information of the object wave is converted into the amplitude variation, and thereby can be stored in the AM-only wave-guiding structure.

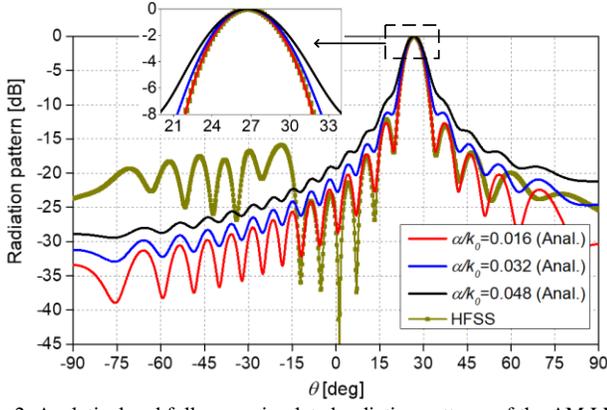

Fig. 2. Analytical and full-wave simulated radiation patterns of the AM LWA with different values of attenuation constant.

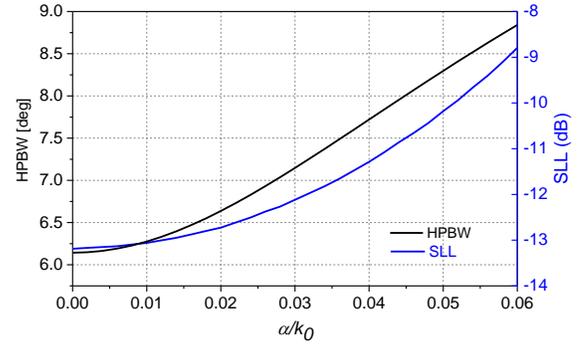

Fig. 3. Calculated HPWB and SLL of the AM LWA versus normalized attenuation constant.

*B. Reconstruction*

To reconstruct the object wave, an identical reference wave should again be utilized to excite the AM wave-guiding structure as shown in Fig. 1(b). However, two considerations should be taken into account for a practical AM LWA. First, the length of an LWA is finite, which causes truncation effect for the reference wave. On the other hand, the aperture amplitude distribution of an LWA is exponentially tapered due to the Ohmic and dielectric losses, and more importantly, the leaky effect. As a result, the reference wave should be modified as following for a practical LWA:

$$E'_r = A_r \exp(-j\beta_{cx}x)\exp(-\alpha x)\Pi(x) \quad (5)$$

where $\alpha$ is the attenuation constant of the LWA, and $\Pi(x)$ is the rectangular function, assuming the excitation source is located at $x = -L/2$ and the power is delivered toward the LWA with a length of $L$:

$$\Pi(x) = \begin{cases} 1, & |x| < L/2 \\ 0, & |x| > L/2 \end{cases} \quad (6)$$

Therefore, the total generated wave on the AM LWA can be written as:

$$\begin{aligned}E_t &= A(x)E'_r \\ &= (A_o^2 + A_r^2)A_r \exp[-j(\beta_{cx}-j\alpha)x]\Pi(x) \\ &\quad + A_r^2 A_o \exp[-j(k_0\sin\theta_b - j\alpha)x]\Pi(x) \\ &\quad + A_r^2 A_o \exp[-j(2\beta_{cx}-k_0\sin\theta_b - j\alpha)x]\Pi(x)\end{aligned} \quad (7)$$

Since a slow-wave transmission is assumed (i.e., $\beta_{cx} > k_0$), the first term on the right-hand side of (7) is a bounded surface wave propagating along the wave-guiding structure. The second term is the desired reconstructed wave containing the object wave. The third term is a conjugated version of the object wave. Since its propagating wavenumber $2\beta_{cx} - k_0\sin\theta_b > 2k_0 - k_0\sin\theta_b > k_0$, the third conjugated term is a non-radiating bounded mode. As a result, although three types of waves are generated by exciting the AM hologram, only the desired reconstructed wave containing the object wave is a fast wave and can be radiated into free space.

Similar to the classic optical Gabor hologram, the AM leaky-wave antenna records the intensity of the interferogram pertinent to both the amplitude and phase of the object wave. But different from the optical Gabor hologram, it is the surface wave rather than the space wave that is used as the reference wave for AM LWAs. Furthermore, Gabor holograms suffer from the limitation of the twin-image problem, i.e., overlapping the desired object wave with the conjugated object wave [25]. Fortunately, this is never a problem for AM LWAs because the conjugated object wave in the third term of (7) is a bounded surface wave which does not contribute to the far field radiation pattern.

The reconstructed wave consists of the object wave multiplied by the exponential taper and the rectangular function as given in (7). The far-field radiation pattern of the AM LWA can be calculated by taking the Fourier transform:

$$\begin{aligned}E(\theta) &= \mathcal{F}[\{A_r^2 A_o \exp[-j(k_o\sin\theta_b - j\alpha)x]\,\Pi(x)\} \\ &= A_r^2 A_0 \delta[\beta_x - (k_0\sin\theta_b - j\alpha)] \otimes \frac{2\sin(\beta_x L/2)}{\beta_x} \\ &= 2A_r^2 A_0 \frac{\sin\{[k_0(\sin\theta - \sin\theta_b) + j\alpha]\cdot L/2\}}{k_0(\sin\theta - \sin\theta_b) + j\alpha}\end{aligned} \quad (8)$$

In (8), the symbol $\otimes$ represents a convolution operator. To investigate the role of the attenuation constant, we consider an AM LWA case where the intended radiation angle of the object wave is $\theta_b = 27^o$ at 55 GHz and it is recorded in a slow-wave AM transmission line with a length $L = 50$ mm. These parameters are adopted in the IDW LWA in Section III. The far-field radiation patterns calculated by (8) for three different values of attenuation constant are depicted in Fig. 2. It can be seen that the positions of the main lobe, side lobe, and dips remain unchanged for different attenuation constants. Nevertheless, higher side lobe levels (SLLs), larger half power beamwidth (HPBW), and more indistinct dips are observed for increasing the attenuation constant. The calculated HPBW and SLL as a function of the attenuation constant are further illustrated in Fig. 3. The HPBW of the LWA increases from 6.19$^o$ to 8.57$^o$ and the SLL increases from -13.3 to -8.3 dB as the normalized attenuation constant increases from 0 to 0.06. A larger HPBW for increasing the attenuation constant is attributed to the reduced effective aperture of the LWA.

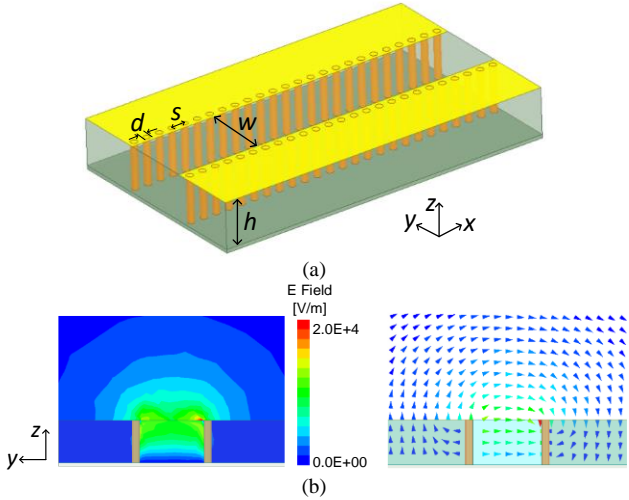

(a)

(b)

Fig. 4. Configuration of the uniform substrate integrated IDW without carrying any information of object wave. Simulated (a) magnitude and (b) vector distributions of the fundamental $TE_{01}$ mode of the substrate integrated IDW in the transversal plane ($yz$-plane).

### C. Unified Holographic Theory and Spatial Spectrum Theory

The concept of AM LWA originates from the amplitude modulation technique in classical communications theory [23]. In this work, the AM LWA is treated as an amplitude-recording hologram based on the classical Gabor holography. The consistency between these two theories is further demonstrated as follows.

For the recording process in Section II-A, the intensity of the interference pattern in (4) can be reorganized as:

$$A(x) = (A_o^2 + A_r^2)[1 + \frac{2A_r A_o}{A_o^2 + A_r^2}\cos(\beta_{cx}x - k_0\sin\theta_b x)]$$
$$= A_c[1 + M\cos(\frac{2\pi}{\Lambda}x)] \quad (9)$$

where $M$ ($0 < M < 1$) is the modulation index, and $\Lambda$ is the periodicity of the sinusoidal modulating wave, with the definitions of

$$A_c = A_o^2 + A_r^2 \quad (10)$$

$$M = \frac{2A_r A_o}{A_o^2 + A_r^2} \quad (11)$$

$$\frac{2\pi}{\Lambda} = \beta_{cx} - k_0\sin\theta_b \quad (12)$$

It can be seen from (9) that the intensity of the interference pattern of the two waves can be written in the form of sinusoidal amplitude modulation – in consistent with the finding in [23] that a sinusoidally AM LWA can generate a high-gain pencil beam.

From the perspective of spatial spectrum theory, the AM LWA is interpreted as a waveform modulator to shift the spatial spectrum of the modulating wave up to the carrier spatial frequency. The free space acts as a spatial band-pass filter which only allows the spatial spectrum fallen within the visible region to radiate. A sinusoidal amplitude variation is required in order to generate the desired spatial frequency impulse. The developed holographic theory in this work provides a completely different physical interpretation of AM LWAs. The reason why a sinusoidal amplitude modulation is required to generate a high-gain beam is due to the fact that the intensity of the interference pattern between the slow reference wave and the object plane wave is a sinusoidal distribution, as depicted in Fig. 1(a).

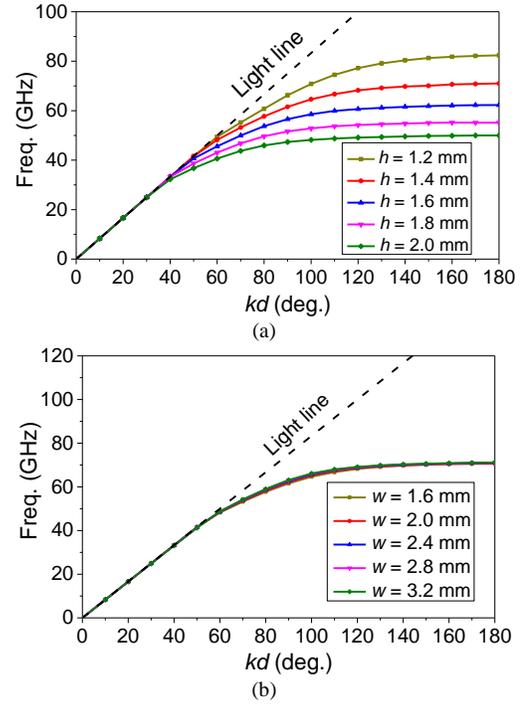

(a)

(b)

Fig. 5. Dispersion diagram of the substrate integrated IDW transmission line with different values of (a) substrate thickness $h$ and (b) aperture width $w$.

## III. FAR-FIELD HOLOGRAPHIC AM LWA

IDW, a rectangular groove filled with dielectric, is an appealing transmission line candidate at mmW frequencies owing to its attractive characteristics, such as low propagation loss and ease of fabrication [26]. Furthermore, substrate integrated IDW, as one type of substrate integrated circuits [27], can be easily implemented using the standard PCB process and integrated with circuits. Many efforts in open literature have been made to convert the bounded slow waves in IDWs into radiating modes, including adding metal strip grating on the air-dielectric interface [28], sinusoidal modulation of the depth of the dielectric groove, and thereby modulating the surface impedance [29], and IDW-excited patch array [30]. The developed AM holographic theory in Section II opens another avenue to realize IDW-based LWAs.

Fig. 4(a) shows the configuration of the uniform substrate integrated IDW transmission line, in which two rows of the metallized via work as the vertical walls to realize the traditional rectangular IDW in planar form. The diameter $d$ and spacing of the metallized via holes $s$ are 0.3 and 0.5 mm, respectively. The aperture width of the dielectric aperture is $w$ and the thickness of the substrate is $h$. The substrate used here is Rogers 5880 with a relative permittivity of 2.2. The fundamental mode of the substrate integrated IDW is $TE_{01}$ mode, whose E-field distribution in the transversal plane is shown in Fig. 4(b). It can be seen that the energy is mainly confined in the substrate and IDW aperture.

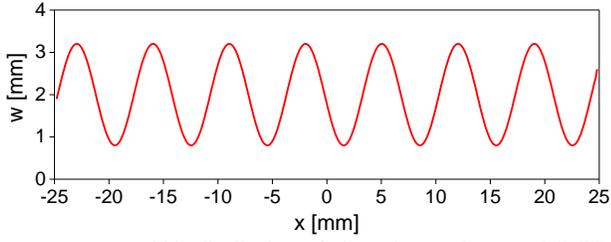

Fig. 6. Aperture width distribution of the substrate integrated IDW for high-gain pencil beam generation.

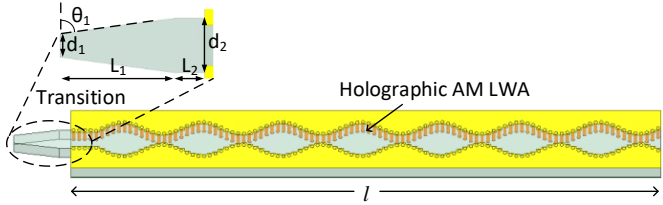

Fig. 7. Configuration of the holographic AM LWA based on substrate integrated IDW for high-gain pencil beam generation. (Dimensions: $d_1 = 0.8$ mm, $d_2 = 1.9$ mm, $L_1 = 4$ mm, $L_2 = 1$ mm, $l = 52$ mm, $\theta_1 = 82°$.)

The dispersion curves of the substrate integrated IDW with different values of dielectric thickness $h$ and aperture width $w$ simulated using the eigenmode solver of ANSYS HFSS are given in Figs. 5(a) and (b), respectively. It can be seen that the dispersion curves are in the slow-wave region and hence the wave is bounded and propagating along the IDW transmission line. Furthermore, the dispersion relation is mainly affected by the dielectric thickness $h$ while insensitive to the aperture width $w$. Nevertheless, the field intensity of the substrate integrated IDW will increase when the aperture width is reduced due to the compressed field in the transversal plane ($yz$-plane). As a result, the amplitude of the guided wave on the substrate integrated IDW can be modulated by simply varying the aperture width $w$ without changing the surface impedance. Note that the dispersion diagrams given in Fig. 5 are just to show that the phase of the carrier wave of the IDW is not modulated, while the dispersion diagram has not been used during the AM LWA synthesis process. As a result, the time-consuming eigenmode simulation can be actually avoided for AM LWAs design.

The AM IDW was designed using the amplitude holography theory introduced in Section II. The thickness of the substrate is 1.4 mm, which consists of two Rogers 5880 substrates with a thickness of 0.508 and 0.787 mm bonded by a 0.1 mm-thick Rogers 4450F bonding film. The normalized propagation wavenumber of the traveling wave in the substrate integrated IDW, i.e., reference wave, is $\beta_{cx}/k_0 = 1.23$ at 55 GHz based on the dispersion curve in Fig. 3. The intended main beam direction of the holographic AM leaky-wave antenna is 27° at 55 GHz. The intensity of the interferogram between the object and reference waves can be determined by (4) with the parameters of $A_r = 1$ and $A_o = 0.34$. The aperture width $w$ of the substrate integrated IDW is tuned accordingly to record the intensity of the interferogram and is shown in Fig. 6.

The configuration of the holographic AM LWA based on the substrate integrated IDW is shown in Fig. 7, which consists of a taper transition part and a sinusoidally modulated substrate integrated IDW transmission line. The holographic AM LWA has a length of $l = 52$ mm, which is excited by a standard rectangular waveguide (WR15). The taper transition, acting as an impedance transformer, is directly inserted into the rectangular waveguide to excite the fundamental $TE_{01}$ mode of the substrate integrated IDW. A waveguide is used as the feed here for testing purpose, and a ground coplanar waveguide (GCPW) can be used to excite the substrate integrated IDW when integrating the holographic AM LWA with mmW circuits.

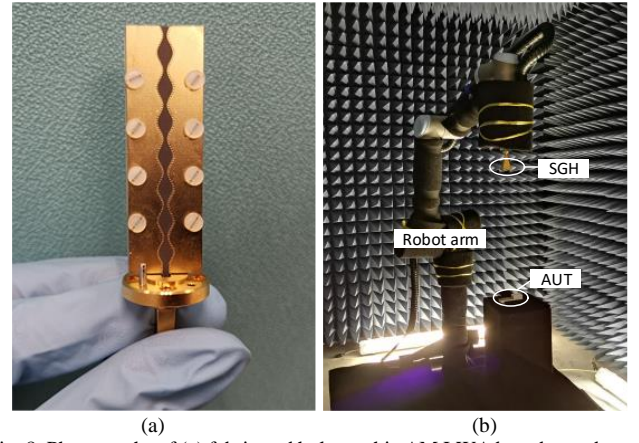

Fig. 8. Photographs of (a) fabricated holographic AM LWA based on substrate integrated IDW and (b) measurement setup for far-field radiation pattern measurement.

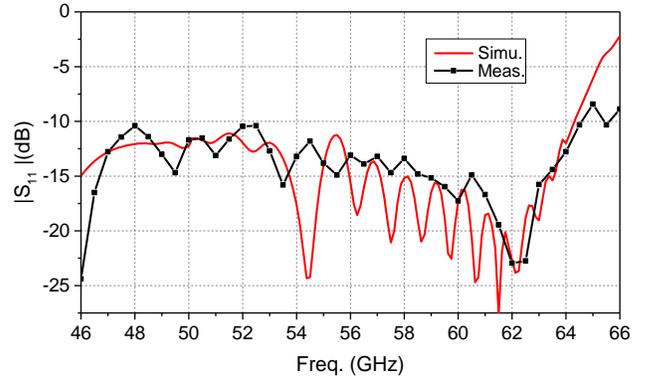

Fig. 9. Simulated and measured $|S_{11}|$ of the holographic AM LWA for high-gain pencil beam applications at different frequencies.

The radiation performance of the holographic AM LWA was modeled and simulated in ANSYS HFSS. The simulated normalized attenuation constant $\alpha/k_o$ is 0.016 based on the simulated $S$-parameter. The simulated far-field radiation pattern at 55 GHz is given in Fig. 2. It can be observed that the analytical result based on the holographic theory agrees well with the simulated one. Specifically, both the analytical and simulated main beam directions $\theta_b$ are 27°. The analytical and simulated HPBW are 6.36° and 6.39°, respectively. The analytical and simulated SLLs are -12.6 and -12.1 dB, respectively. The positions of the sidelobe and dips are also in good agreement, demonstrating the effectiveness of the developed amplitude holography theory in this work. Some higher SLLs in backward direction for the full-wave simulation result may be caused by the diffraction and reflection of the residue power at the end of LWA, which are not considered in the amplitude holography theory.

Fig. 8(a) shows the manufactured holographic AM LWA prototype, fabricated by the standard PCB process. The

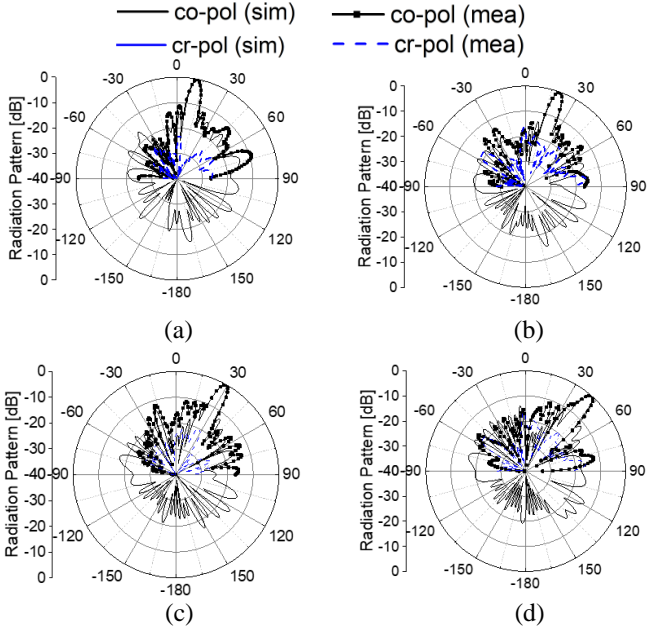

TABLE I
COMPARISONS WITH OTHER *V*-BAND FAR-FIELD LWAs

| Ref. | Freq. (GHz) | Antenna type | Gain (dBi) | Scan range | SLL (dB) | Radiation efficiency |
|---|---|---|---|---|---|---|
| [32] | 52-63 | MEMS LWA | 11.4 | 9° (3°-12°) | ~-8 | NA |
| [33] | 56-64 | MEMS LWA | 15.4 | 8° (41°-49°) | -8 | NA |
| [34] | 58-67 | LTCC SIIG | 11.7 | 21° (7°-38°) | -7 | 90% |
| [29] | 50-85 | Bulky 3D printed IDW | 14.2 | 49° (-9°-40°) | ~-6 | 75% |
| [35] | 55-65 | Single-layer PCB CRLH SIW | 14.5 | 120° (-72°-48°) | ~-10 | 71% |
| This work | 47-60 | Single-layer PCB IDW | 15.0 | 33° (9°-42°) | -10 | 73% |

MEMS refers to micro-electro-mechanical systems.
LTCC refers to low temperature co-fired ceramic.
SIIG refers to substrate integrated image guide.
CRLH refers to composite right/left-handed.
SIW refers to substrate integrated waveguide.

Fig. 10. Simulated and measured normalized H-plane (*xz*-plane) radiation patterns of the holographic AM LWA for far-field high-gain applications. (a) 48, (b) 52, (c) 56, and (d) 60 GHz.

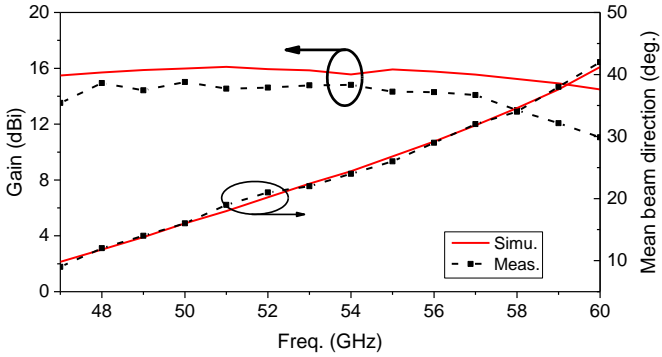

Fig. 11. Simulated and measured gains and main beam direction of the holographic AM LWA versus frequency.

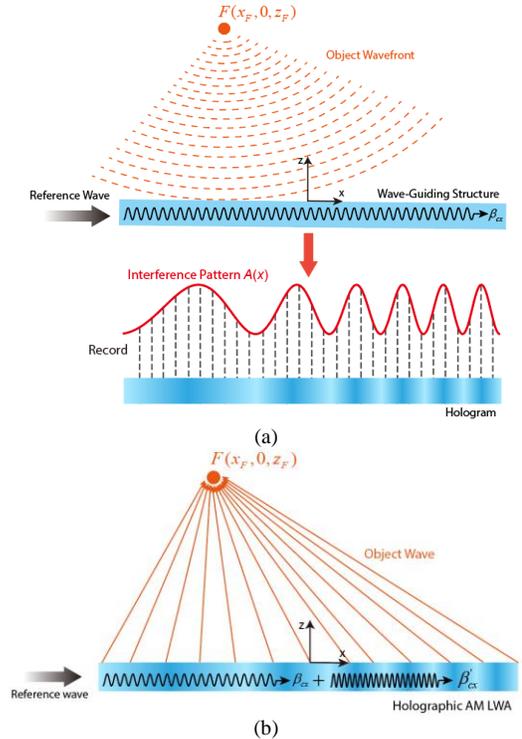

Fig. 12 Concept of the holographic AM LWA for NFF applications: (a) Recording process; (b) Reconstruction process.

sinusoidally AM leaky-wave antenna was measured in an in-house far-field/near-field reconfigurable robotic mmW antenna measurement system [31], as shown in Fig. 6(b). The antenna under test (AUT) is deployed as the transmitting antenna while a standard gain horn (SGH) (QWH-VPRR00) is used as the receiving antenna. The *S*-parameter of the fabricated substrate integrated IDW was measured by a vector network analyzer (E8361A PAN Network Analyzer). The full-wave simulated and measured results of the $|S_{11}|$ are shown in Fig. 9. The $|S_{11}|$ of the holographic AM LWA is smaller than -10 dB from 46 to 64 GHz.

The simulated and measured normalized H-plane (*xz*-plane) radiation patterns at different frequencies are illustrated in Fig. 10. It can be observed that the pencil beam can be successfully generated by the holographic AM LWA. Good agreement between the simulated and measured results can be observed. The measured cross-polarization levels are below -20 dB within the whole frequency band. Fig. 11 shows the simulated and measured gains of the holographic AM LWA versus frequency. The measured gain varies among 11−15 dBi from 47 to 60 GHz. The measured gain is lower than the simulated one by about 1.5 dB on average, which is primarily caused by the increased insertion loss of the substrate in the mmW band and the fabrication error. The simulated and measured main beam angles are also given in Fig. 11. The measured main beam angles are 26.5° at 55 GHz, agreeing well with the desired object wave direction of 27°. Both simulated and measured results show that the main beam direction of the holographic AM LWA can be scanned from 9° to 42° as the frequency varies from 47 to 60 GHz. Table I summarizes the comparison with some previously reported 1-D *V*-band LWAs. It can be seen that the designed far-field IDW LWA has a higher gain and wider scanning range compared to most of the reported *V*-band LWAs except the composite right-/left-handed (CRLH) LWA in [35], where a complicated optimization approach based on full-wave simulation is utilized to optimize the LWA structure. Furthermore, the developed AM IDW-based LWA can be easily implemented using a single-layer PCB process and integrated with circuits, making it is an appealing high-gain antenna for mmW applications.

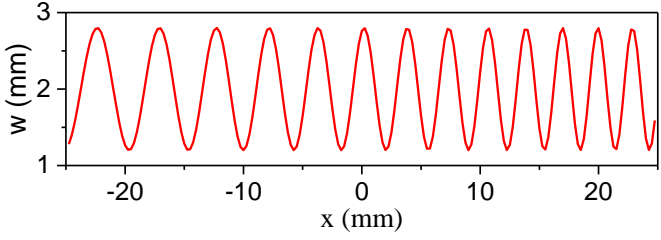

Fig. 13. Aperture width distribution of the substrate integrated IDW for NFF applications.

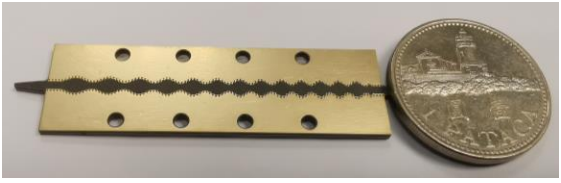

Fig. 14. Photograph of the fabricated holographic AM LWA based on substrate integrated IDW for NFF applications.

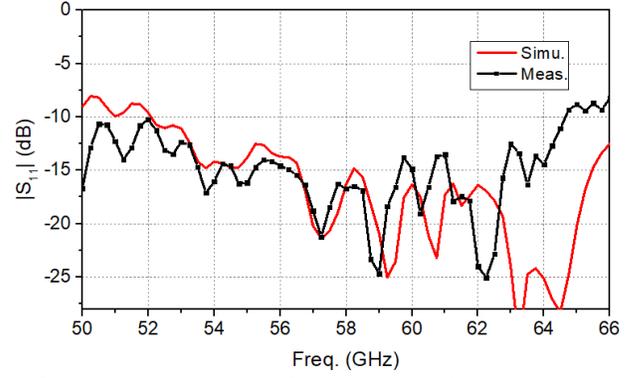

Fig. 15. Simulated and measured $|S_{11}|$ of the holographic AM LWA for NFF applications at different frequencies.

## IV. NEAR-FIELD FOCUSING HOLOGRAPHIC AM LWA

NFF antenna, capable of concentrating electromagnetic power into a small spot, is highly demanded in various applications, including noncontact sensing, RFID system, mmW imaging, and wireless transmission energy systems [36]-[37]. Although both holographic theory and spatial frequency spectrum analysis can be used to design far-field AM LWAs as demonstrated in Section III and in [23], the holographic theory in this paper provides a more intuitive approach to synthesize AM LWAs for NFF applications.

For the NFF hologram, the object wave is a spherical wave converging to the focus point $F = (x_F, 0, z_F)$, as shown in Fig. 12(a). The object wave can be expressed as:

$$E_o = A_o \exp(jk_0\sqrt{(x-x_F)^2 + (z-z_F)^2}) \quad (13)$$

The unmodulated slow traveling wave described by (2) is again used as the reference wave. The amplitude-only wave-guiding hologram records the intensity of the interferogram between the object and reference waves, which can be obtained by substituting (2) and (13) into (3). The results can be written as:

$$A(x) = A_o^2 + A_r^2 + 2A_rA_o\cos[\beta_{cx}x + k_0\sqrt{(x-x_F)^2 + z_F^2}] \quad (14)$$

For verification purpose, assume that the substrate integrated IDW is used as the AM hologram to focus the energy at the focal point $F$ ($x_F = -10$ mm, $y_F = 0$, $z_F = 40$ mm) at 55 GHz. The calculated result of the aperture width of the substrate integrated IDW is shown in Fig. 13 with the parameters of $A_r = 1$, $A_o = 0.21$. A larger modulation period at the beginning can be seen and it gradually decreases along the substrate integrated IDW transmission line.

TABLE II
COMPARISONS WITH OTHER REPORTED NEAR-FIELD FOCUSING LWAS

| Ref. | Freq. (GHz) | Antenna type | Planar struc. | FoV | Spot size | NF SLL (dB) | Scan loss (dB) |
|---|---|---|---|---|---|---|---|
| [38] | 18-20 | Corrugated waveguide | No | ~5.7° | 0.65$\lambda_0$ | NA | NA |
| [39] | 33-39 | SIW | No | NA | NA | ~-7.5 | 4.5 |
| [40] | 14-16 | Microstirp line | Yes | ~15.7° | 1$\lambda_0$ | NA | NA |
| [41] | 33-37 | SIW | Yes | ~24.8° | NA | ~-9 | 5 |
| This work | 51-59 | Single-layer IDW | Yes | 26.2° | 0.82$\lambda_0$ | -10 | 2.1 |

For the reconstruction process, the reference wave $E'_r$ in (5) is utilized to excite the amplitude-only hologram, as shown in Fig. 12(b). The overall reconstructed wave can be written as:

$$\begin{aligned}E_t &= A(x)E'_r \\ &= (A_o^2 + A_r^2)A_r\exp[-j(\beta_{cx} - j\alpha)x]\Pi(x) \\ &\quad + A_r^2A_0\exp(jk_0\sqrt{(x-x_F)^2 + z_F^2})\exp(-\alpha x)\Pi(x) \\ &\quad + A_r^2A_0\exp[-j(2\beta_{cx}x + k_0\sqrt{(x-x_F)^2 + z_F^2})]\exp(-\alpha x)\Pi(x)\end{aligned} \quad (15)$$

Again, only the second term on the right-hand side of (15) contains the desired reconstructed object wave converging to the focal point, while the first and third terms are residual slow waves, propagating along the wave-guiding structure as illustrated in Fig. 12(b).

The photograph of the fabricated substrate integrated IDW-based LWA is shown in Fig. 14. An identical taper transition to that of the previous sinusoidal AM LWA is added between the holographic AM LWA and feeding waveguide for impedance matching purpose. Fig. 15 plots the simulated and measured $|S_{11}|$ of the NFF AM LWA. It can be seen that the measured $|S_{11}|$ is better than -10 dB from 50 to 64.5 GHz. Fig. 16 plots the simulated and measured near-field power intensities in the xz-plane at different frequencies. The simulated and measured results agree reasonably well. Some ripples in the measured results can be observed from Fig. 16, which is primarily caused by the multiple reflections between the metallic receiving probe and the antenna under test in the near-field measurement process. It can be seen from Fig. 16 that the reconstructed electromagnetic wave from the holographic AM LWA can be concentrated into an ellipsoidal focal region around the desired focal point ($x_F = -10$ mm, $y_F = 0$, $z_F = 40$ mm) at 55 GHz. As the operating frequency increases, the focal point also scans from the backward towards forward directions due to the dispersion of the substrate integrated IDW. The measured Field of View (FoV) of the IDW LWA is 26.2º (from 0º to 26.2º). Fig. 17 shows the simulated and measured power intensities along the line $z = 40$ mm at different frequencies. Good agreement between the simulation and measurement in terms of spot size and near-field (NF) SLL can be observed. The measured -3dB spot size of the AM LWA at 55 GHz is 4.47 mm, corresponding to $0.82\lambda_0$. The measured NF SLL is -10 dB at 55 GHz. The measured scan loss, defined as the discrepancy of electric field intensity at the focal point when scanning [42], is 2.1 dB as frequency varies from 51 to 59 GHz. Table II shows the comparison between the proposed AM NFF LWA with previously reported NFF LWAs. It can be observed that the proposed IDW-based LWA has advantages of better near-field frequency scanning performance in terms of larger FoV, lower NF SLL within the frequency band, and lower scan loss.

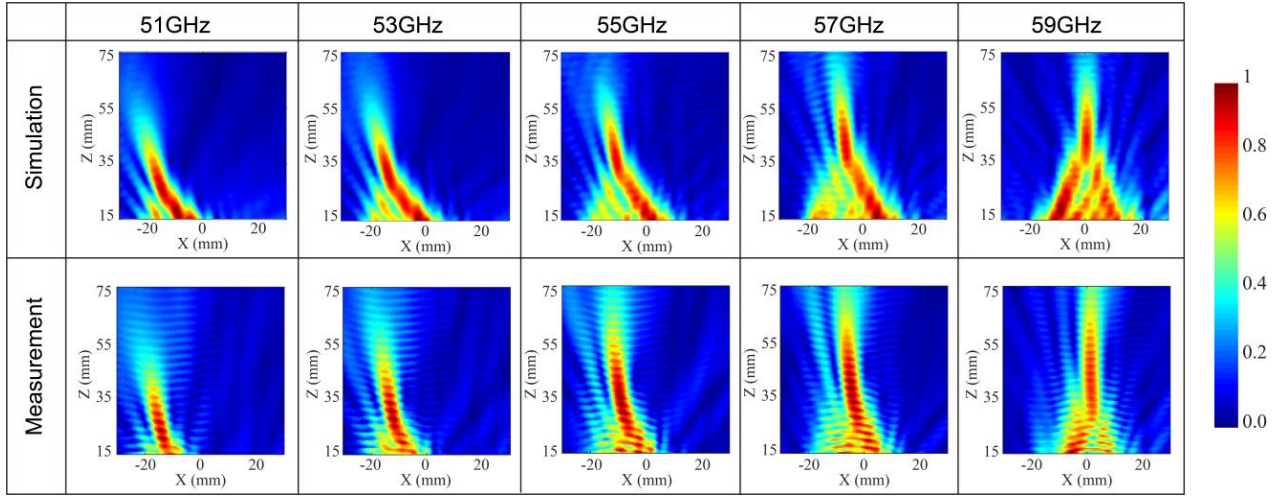

Fig. 16. Simulated and measured near-field power intensities in the *xz*-plane at different frequencies.

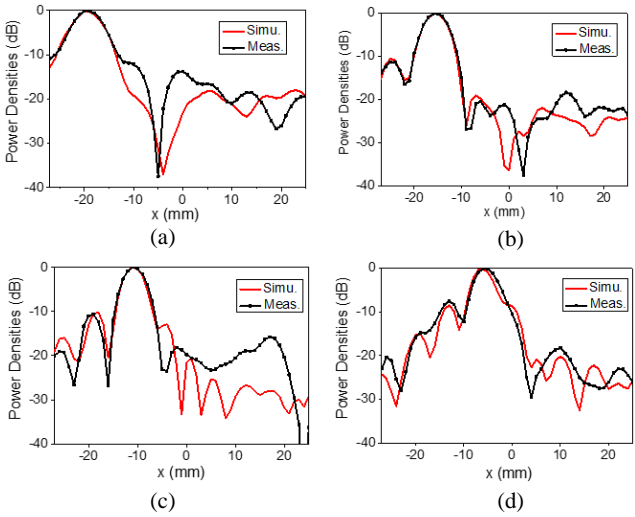

Fig. 17. Simulated and measured normalized power densities along the line $z = 40$ mm. (a) 51, (b) 53, (c) 55, and (d) 57 GHz.

## V. CONCLUSION

The concept of AM LWA originates from classical communications theory, while in this paper, we use another classical theory, i.e., holography technique in the optical domain, to reinterpret this promising concept. The AM holographic theory provides an intuitive and insightful explanation as to why the amplitude-only modulation of the traveling wave can carry both the amplitude and phase information of the object wave. In the first demonstration case, a sinusoidal AM LWA based on the substrate integrated IDW is designed for far-field high-gain pencil beam applications. The measured main-beam direction of the holographic AM LWA scans from 16º to 42º as the operating frequency varies from 50 to 60 GHz. In the second case, the object wave is a spherical wave and the corresponding holographic AM LWA is synthesized for NFF applications. Both simulated and measured results show that the holographic AM LWA can concentrate the reconstructed wave into a desired near-field focal spot.